\documentclass[aps,prb,twocolumn,superscriptaddress]{revtex4-2}

\usepackage{graphicx}
\usepackage{dcolumn}
\usepackage{bm}
\usepackage{amsmath}
\usepackage{amssymb}
\usepackage{cases}

\usepackage{color}

\newcommand{\He}{$^4$He}
\newcommand{\et}{$\it{et \textrm{ } al. }$ }
\newcommand{\ie}{$\it{i. e.}$}
\newcommand{\dd}{\textrm{d}}
\newcommand{\B}[1]{\bm{#1}}
\newcommand{\SI}[1]{\textcolor{black}{#1}}

\begin{document}


\title{Bathtub vortex in superfluid $^4$He}



\author{Sosuke Inui}
\affiliation{Department of Physics, Osaka City University, 3-3-138 Sugimoto, 558-8585 Osaka, Japan}
\author{Tomo Nakagawa}
\affiliation{Department of Physics, Osaka City University, 3-3-138 Sugimoto, 558-8585 Osaka, Japan}
\author{Makoto Tsubota}
\affiliation{Department of Physics \& Nambu Yoichiro Institute of Theoretical and Experimental Physics (NITEP) \& The OCU Advanced Research Institute for Natural Science and Technology (OCARINA), Osaka City University, 3-3-138 Sugimoto, 558-8585 Osaka, Japan}

\date{\today}

\begin{abstract}
We have investigated the structure of macroscopic suction flows in superfluid \He{}.
In this study, we primarily analyze the structure of the quantized vortex bundle that appears to play an important role in such systems.
Our study is motivated by a series of recent experiments conducted by a research group in Osaka City University [Yano \et{}, J. Phys. Conf. Ser. $\B{969}$, 012002 (2018)]; they created a suction vortex using a rotor in superfluid \He{}.
They also reported that up to $10^4$ quantized vortices accumulated in the central region of the rotating flow.
 The quantized vortices in such macroscopic flows are assumed to form a bundle structure; however, the mechanism has not yet been fully investigated.
Therefore, we prescribe a macroscopic suction flow to the normal fluid and discuss the evolution of a giant vortex (\ie{}, one with a circulation quantum number exceeding unity) and a bundle of singly quantized vortices from a small number of seed vortices.
Then, using numerical simulations, we discuss several possible characteristic structures of the bundle in such a flow, and we suggest that the actual steady-state bundle structure in the experiment can be verified by measuring the diffusion constant of the vortex bundle after the macroscopic normal flow has been switched off.
By applying extensive knowledge of the superfluid \He{} system, we elucidate a new type of macroscopic superfluid flow and identify a novel structure of quantized vortices.  

\end{abstract}

\pacs{xxxx}

\maketitle


\section{Introduction}
%
We often encounter ``vortices'' of various length scales: the dropping of milk into coffee, whirlpools, the Great Red Spot on the surface of Jupiter, and so on.
The suction vortex, also referred to as the ``bathtub vortex,'' is one of the most familiar classical vortices; it can be easily produced by unplugging a bathtub filled with water.
However, this vortex’s simple generation procedure does not entail that its structure can be easily understood.
Indeed, despite several attempts, no theory of the vortex has yet been completed \cite{Lundgren85, Andersen06, Mulligan18}.
In this paper, we elucidate the bathtub vortex from a different perspective: that of a bathtub vortex in superfluid \He{}.

Liquid \He{}, at a saturated evaporation pressure below the lambda point $T_\lambda \approx 2.17$ K, exhibits superfluidity; in this state, its sheer viscosity vanishes and a number of eccentric phenomena (e.g., fountain and capillary effects) can be observed.
These effects are often explained using a phenomenological model (the so-called ``two-fluid model’’ \cite{Tisza38,Landau41,Tilley90}), in which the superfluid \He{} at $ 0 <  T < T_\lambda $ features two fluid components: an inviscid superfluid with density $\rho_s(T)$ and a viscous normal-fluid with density $\rho_n(T)$.
One of the most notable properties of superfluids is that their circulation $\kappa \equiv \oint_\mathcal{L} \B{v} \cdot d \B{l}$ can be quantized as 
\begin{equation}
\kappa = \frac{h}{m}n, 
\end{equation}
 where $n$ is an integer, $h$ is Planck's constant, and $m$ is the mass of a \He{} atom. This quantization assumes that the path $\mathcal{L}$ encloses a filamentary topological defect in the superfluid.
The topological defects with a quantized circulation always form closed loops or terminate their ends at boundaries, and thus they are called quantized vortex loops or lines \cite{Donnelly91}.
In a bulk superfluid, the kinetic energy per unit length of the vortex line $\epsilon$ is proportional to $n^2$; thus, it is more energetically stable to have two vortices with $n=1$ than one vortex with $n=2$.
The superfluid system is very clean and offers an ideal experimental environment for many fields of physics; thus, it has been extensively studied over the decades by researchers hoping to understand various physical phenomena, including turbulence \cite{Vinen02,Tsubota09,Tsubota13,Barenghi14}, the Kibble--Zurek mechanism \cite{Zurek85,Bauerle,Bunkov}, and pulsar glitches in neutron stars \cite{Andersson12,Page14}.

\begin{figure}[t!] 
	\includegraphics [width=1\columnwidth]{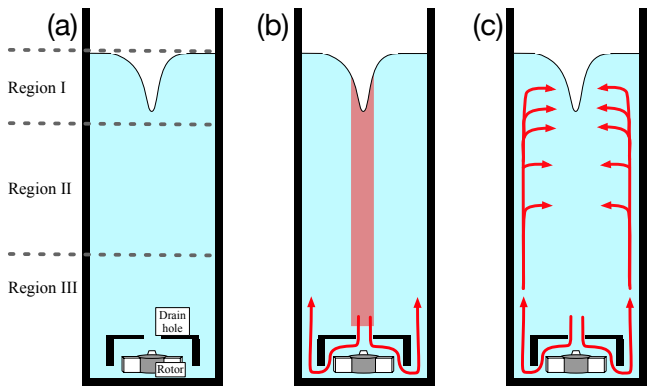}
	\caption{(a) -- Schematic overview of the ``bathtub vortex'' \cite{Yano18,Matsumura19}. The entire length scale of the system (from the surface to the bottom of the fluid) is approximately $20$--$30$ cm. The system can be roughly separated into three regions: Region I, in which the surface of the superfluid \He{} dimples and a large vortex with circulation quantum number $n > 1$ is expected; Region II, in which a steady vortex bundle is thought to develop; and Region III, in which the geometry of the bundle experiences the effects of the bottom boundary. In classical theory, it is understood that an Ekman boundary layer exists at the bottom and an up-flow operates just outside the down-flow travelling through the drain hole. (b) -- Expected normal flow pattern. The rotor repels the fluid and generates a pressure difference. Then, the fluid is forced to flow through the drain hole. The classical theory implies that the down-flow of the normal fluid is tightly confined in the central (shaded) region. (c) -- Expected trajectory of (remnant) quantized vortices. For the system to have a giant vortex, it must be provided with vortices/vorticity externally; otherwise, it does not conserve angular momentum.}
    \label{fig: Schematics}
\end{figure}

In the experiments conducted by a research group at Osaka City University (OCU), Yano \et{} created a macroscopic bathtub vortex by sucking superfluid \He{} (temperature: $T = 1.6$ K) out of a cylindrical container via a drain hole at the bottom, using a rotor (see the schematic overview in Fig. \ref{fig: Schematics} and the figures given in Refs. \cite{Yano18,Matsumura19}). 
The rotor rotated below the drain hole and induced a pressure difference; this allowed normal- and super-fluids to flow.
\SI{When the fluid achieved a steady state, it mimicked the flow of a classical fluid \cite{Yano18} (as it does under steady solid-body rotation by forming a vortex lattice \cite{Osborne50,Andronikashvili66,Madison00,Madison01,Tsubota02,Kasamatsu03}). 
The experimental observations of the shape of the dimpled surface indicates that normal- and super- fluids co-flow with an azimuthal velocity inversely proportional to the radial distance $r$.
 }
The normal fluid has a viscosity; thus, it can be reasonably assumed that its steady-state flow profile resembles the profile discussed in Refs. \cite{Lundgren85, Andersen06}; that is, the down-flow is narrowly confined in the central region, forming a flow tube above the drain hole.
 It is classically understood that an up-flow surrounds the down-flow, owing to the vorticity generated near the central region \cite{Andersen06}.
However, the vorticity of the superfluid is only carried by quantized vortices; therefore, this might not apply in the non-classical case, and the formation of classical-like macroscopic suction flows is not trivial.
Moreover, from observations of second sound attenuation, the vortex line density $L_L$ at the core region (radius $\sim 2$ mm) \SI{well bellow the dimpled surface}  is reported to be as much as $1.3 \times 10^{12}$ m$^{-2}$ \cite{Matsumura19}.
\SI{These vortex lines are thought to be attracted toward the axis of rotation, thereby forming a vortex bundle \cite{Alamri08} through the particular macroscopic flow geometry of the system of two fluids.
In the presence of the downflow of normal fluid, we argue that such a highly accumulated vortex line density in the core region can be developed with a structural pattern inherited from the flow geometry.
Throughout this analysis, we prescribe the profile of the normal fluid velocity and perform a series of numerical simulations to follow the dynamics of individual vortices, rather than a coarse-grained vortex line density field, to investigate the large-scale structure of the vortex bundle.
}
%
%
%

The generation mechanism of a macroscopic bathtub vortex in a superfluid is not trivial.
To understand such novel macroscopic flows in superfluid \He{}, it is necessary to construct models that do not contradict the experimental results; for this, we apply extensive background knowledge on superfluidity and computational techniques developed over several decades.
\SI{ The objective of this study is to qualitatively understand the structure of quantized vortices in such a macroscopic suction flow.
In this article, we argue that 1) the deformed superfluid surface is identified as a giant vortex,  2) a strongly polarized vortex bundle is developed along the rotational axis beneath the dimpled surface, and 3) the polarization of the bundle may be assessed experimentally by measuring the diffusion constant of the bundle.
We divide the system into three regions, as shown in Fig. \ref{fig: Schematics}, based on the boundary conditions.
Region I is where the surface boundary cannot be neglected.
The dimpled surface created in Region I may be identified as a giant vortex from the $1/r$ velocity profile around it (see Sec. \ref{sec: III} for the discussion). 
Region II is a bulk, where there is presumably no giant vortex, but a bundle of singly quantized vortices that would resemble the configuration in Fig. \ref{fig: Diffusion} ($b$) or ($c$).
Region III is a region in which the bottom boundary condition is not negligible.
The flow geometry near the bottom layer, known as an Ekman layer in classical hydrodynamics, is not trivial, and the discussion on the vortex dynamics in this region is beyond the scope of our current work.
 }
\SI{In Sec. \ref{sec: E of M for V} we briefly review the numerical model used to simulate vortex dynamics, \ie{} 3D vortex filament model (VFM)}.
In Sec. \ref{sec: III}, the process of giant vortex production is discussed.
Then, we discuss how vortices are transported from Region I to Region II, using VFM simulations.
In Sec. \ref{sec: IV}, we show that, depending on the geometry of the normal fluid flow, two characteristic vortex bundle structures are possible in Region II: a linear-vortex structure and a cylindrical vortex-layer-like structure.
In Sec. \ref{sec: V}, \SI{we argue that the large scale vortex bundle structure may be determined by the experimental observation of the diffusion constant of the vortex bundle.
We qualitatively estimate the characteristic diffusion time scale of the bundle from the VFM simulations. 
}
Finally, in Sec. \ref{sec: VI}, we summarize the overall structure of a bathtub vortex.

\section{Equation of Motion for Vortices} \label{sec: E of M for V}
%
The core radius of a quantized vortex in superfluid \He{} is of the order of $\AA$, and a vortex segment carries a potential flow of velocity $v_s \propto 1/r$ around it, where $r$ is the radial distance from the vortex core.
Thus, quantized vortices are often treated as having a delta-function-like vorticity at position $\B{s}(\xi)$, using the arc length parameterization $\xi$.
Thus, the motion of a quantized vortex obeys the Helmholtz's theorems and follows the local superfluid flow $\B{v}_{s}(\xi)$.
However, at finite temperatures, the temperature-dependent mutual friction terms $\alpha$ and $\alpha '$ become significant, and the equation of motion is \cite{Schwarz85}
\begin{equation} \label{eq: EoM VFM}
\begin{split}
\frac{\dd \B{s}(\xi,t)}{\dd t} = \B{v}_\textrm{s} & + \alpha \B{s}^\prime(\xi)  \times (\B{v}_\textrm{n} - \B{v}_\textrm{s} )\\
&- \alpha^\prime \B{s}^\prime (\xi)  \times  \left[ \B{s}^\prime (\xi) \times (\B{v}_\textrm{n} - \B{v}_\textrm{s} ) \right] ,\\
\end{split}
\end{equation}
where $\B{v}_s$ and $\B{v}_n$ are the velocity fields of super- and normal-fluids, respectively; the prime symbol $'$ denotes the derivative with respect to arc length $\xi$. 
\SI{Therefore, we can calculate the time-evolution of a vortex, once we obtain the velocities, $\B{v}_s$ and $\B{v}_n$, at $\B{s}(\xi)$. }
%
%

\SI{In Region II, we consider a symmetrically rotating flow of normal fluid along the $z$-axis that resembles a Rankine vortex velocity profile of the form
}
\begin{equation} \label{eq: vn in}
\B{v}_\textrm{n}(r,\phi,z) = \left(
\begin{array}{c} 
0\\
\frac{\Gamma_n}{2\pi}\frac{r}{R_0^2} \\
\SI{v_z(r)}
\end{array}
\right)  \quad \text{for $r<R_0$}
\end{equation}
\begin{equation} \label{eq: vn out}
\B{v}_\textrm{n}(r,\phi,z) = \left(
\begin{array}{c}
0\\
\frac{\Gamma_n}{2\pi}\frac{1}{r} \\
\SI{v_z(r)}
\end{array}
\right)  \quad \text{for $r>R_0$},
\end{equation}
where $R_0$ is the radius of the down-flow tube (which is the same size as the drain hole at the bottom of the container), and $\Gamma_n$ is the circulation of the normal fluid.
\SI{The vertical velocity profile $v_z(r)$ is not known, experimentally nor theoretically.
 Since we qualitatively investigate the macroscopic structure that is imprinted on a vortex bundle by such a flow in Region II, we assume that the structure is not highly dependent on the detail of the flow profile $v_z(r)$.
Thus, for simplicity, we take it as constant if $r<R_0$, and $0$ otherwise, and
To identify the velocity $\B{v}_s$ at $\B{s}(\xi)$ we apply the vortex filament model (VFM), which is briefly explained in the following subsection.}

\subsection{Numerical Method: Vortex Filament Model} \label{sec: II A}
%
First, we consider a 3D vortex line configuration, discretizing it into segments of length $d \xi$.
A vortex segment at $\B{s}(\xi)$ tends to move with velocity $\B{v_s}(\B{s}(\xi))$.
The term $\B{v_s}(\B{s}(\xi))$ can be decomposed into three contributions: the velocity $\B{v}_{s,0}$, which is induced by all vortices in the system; the velocity $\B{v}_{s,\textrm{ext}}$, which is imposed externally; and the velocity $\B{v}_{s,\textrm{b}}$, which is induced by the boundaries.
The superfluid velocity $\B{v}_{s,0}$ at $\xi$ is obtained by calculating the following Biot--Savart integral:
\begin{equation} \label{eq: BS integral}
\begin{split}
\B{v}_{s,0}(\xi) =& \frac{\kappa}{4\pi} \int_{\mathcal{L}} \frac{ \B{s}^{\prime} (\xi_1) \times ( \B{s} (\xi) - \B{s} (\xi_1) ) }{ | \B{s} (\xi) -\B{s}(\xi_1)|^3 } \dd \xi_1   \\
 = & \B{v}_{s,\textrm{loc}} + \B{v}_{s,\textrm{non-loc}}.  \\
\end{split}
\end{equation}
The integral \eqref{eq: BS integral} diverges as $\xi_1 \rightarrow \xi$, because we neglect the core radius $a$ of the vortex.
Computationally, we avoid the divergence by separating out the local term from the total integration path $\mathcal{L}$, to obtain $\B{v}_{s,\textrm{loc}}$ and  $\B{v}_{s,\textrm{non-loc}}$.
Applying the local induction approximation, $\B{v}_{s,\textrm{loc}}$ can be evaluated as $\B{v}_{s,\textrm{loc}} \approx \beta \B{s}^\prime \times \B{s}^{\prime\prime}$, where $\beta = (\kappa/4\pi) \ln(R/a)$.
To solve Eq. \eqref{eq: EoM VFM} and perform the simulation, the path $\mathcal{L}$ is divided into segments of $\Delta \xi$, and the integration in Eq. \eqref{eq: BS integral} is calculated for each segment and at every time step $\Delta t$ in the fourth-order Runge--Kutta scheme.

\section{Giant Vortex and Vortex Transport in Region I} \label{sec: III}
%
\begin{figure}[b!] 
	\includegraphics [width=1\columnwidth]{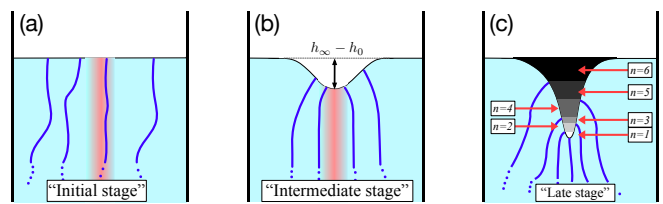}
	\caption{(a) -- (c) Snapshots of the three stages of giant vortex production (color online). The blue lines in each panel represent the singly quantized vortices, and the shaded region around the vertical axis ($z$-axis) represents the region in which the vorticity of the normal fluid accumulates and forms a strong down-flow. Each stage is briefly described as follows: (a) -- Initial stage: vortex lines gather and tend to form a lattice. (b) -- Intermediate stage: the surface of the central region dimples owing to the azimuthal velocity, which is inversely proportional to the radial distance $r$ and pressure difference. (c) -- Late stage: the dimple grows to become a cavity by ``absorbing'' singly quantized vortices. At this stage, the normal fluid circulation $\Gamma_n$ is not necessarily equal to that of the giant vortex, $\kappa n_\textrm{giant}$.}
    \label{fig: Giant}
\end{figure}
%
\begin{figure*}[t!] 
	\includegraphics [width=2\columnwidth]{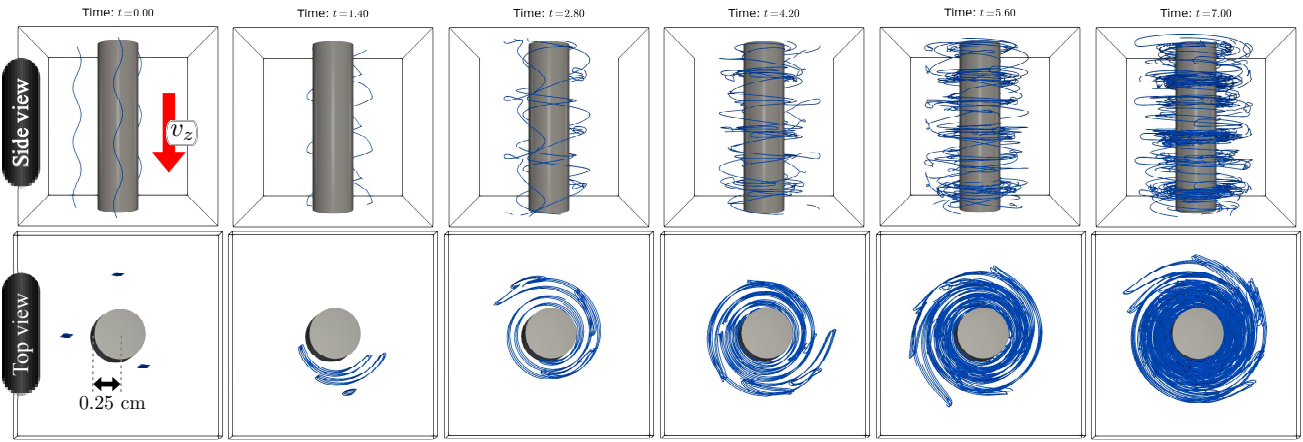}
	\caption{Series of snapshots of VFM simulation at $t = 0.0$, $1.4$, $2.8$, $4.2$, $5.6$, and $7.0$ s from left to right, respectively (color online). A box measuring $2$ cm in each dimension is drawn for reference. The cylinder (radius: $0.25$ cm) in each panel represents a giant vortex, around which the circulation of both fluids are non-zero. In the system, the external normal fluid velocity $v_z$ is applied downward. The top and bottom surfaces of the box are subject to the periodic boundary condition. }
    \label{fig: Region1_VFM}
\end{figure*}
%
The steady bathtub vortex in \He{} features a deep cavity in the central region.
The shape of the cavity indicates that the azimuthal velocities of both the super- and normal-fluids are inversely proportional to the radial distance $r$ around it.
This implies that, for a fully developed bathtub vortex, the cavity behaves like a giant vortex; that is, a quantized vortex with a circulation quantum number $n > 1$.
%

%
Here, we consider how the giant vortex grows.
One of the most conceivable scenarios of giant vortex production in the initial stages of bathtub vortex evolution is as follows:
First, the vorticity of the normal fluid accumulates in the central region along the $z$-axis, and the quantized vortices are also transported toward the central region from the surrounding bulk fluid.
As these gather, they start to exhibit a collective rotational motion, forming some type of lattice structure; this is analogous to the triangular-lattice formation observed in solid-body rotating superfluid helium \cite{Williams80}, BEC \cite{Madison00,Madison01,Tsubota02,Kasamatsu03}, and superconducting currents \cite{Tinkham}.
Then, the surface of the superfluid \He{} gradually starts to deform in the central region, due to the pressure difference and down-flow.
The surface becomes increasingly deformed and generates a cavity of depth $h_\infty -h_0$ (as measured from the height of the stationary surface $h_\infty$ at $r \rightarrow \infty$) as the vorticity of the normal fluid accumulates and vertical vortices enter the vicinity; we can identify this as a giant vortex of circulation quantum number $n_\textrm{giant}  > 1$.
Taking the cavity depth $h$ as a function of radial distance $r$, $h(0) = h_0$ and $\lim_{r \rightarrow \infty} h(r) = h_{\infty}$; thus, the quantum number $n_\textrm{giant}$ at $h(r)$ can be identified as the number of singly quantized vortex lines attached below the surface, as shown in Fig. \ref{fig: Giant} (c).
In a steady state, the macroscopic flow profiles of the super- and normal-fluids coincide with each other, to minimize the mutual friction; this means that the circulation of each fluid around the entire system is equal; that is, $\Gamma_n = \Gamma_s + \kappa N_\textrm{vor}$.
Here, $\Gamma_s = \kappa n_\textrm{giant} $ with $\kappa = h/m$, and $N_\textrm{vor}$ is the number of freely floating vortex lines.

%
If the system is ideally clean (\ie{}, no remnant vortex rings exist), then after a sufficiently long time, $\Gamma_n = \Gamma_s$ and $\kappa N_\textrm{vor} = 0$ are satisfied, because all the singly quantized vortices are ``absorbed'' into the giant one.
However, because of the geometry of the experimental setup, vortex rings can be constantly transported to the central regions from the side, under the macroscopic flow generated by the rotor (see Fig. \ref{fig: Schematics} (c) ).
We conducted numerical simulations to qualitatively assess the vortex line distributions in the presence of flows proportional to $1/r$; that is, the azimuthal velocity profiles for normal- and super-fluids were $v_n = \Gamma_n/2\pi r$ and $v_s = \Gamma_s/2\pi r$, respectively, for an $r$ outside the giant vortex (radius: $0.25$ cm).
%

%
We consider the case in which the normal fluid velocity is steady, but the giant vortex of the superfluid is still growing; that is, $\Gamma_n >\Gamma_s$.
Figure \ref{fig: Region1_VFM} shows a series of snapshots of the simulation, conducted with the parameters $\Gamma_n = 5.0 \times 10^{-4}$ m$^2$/s  and  $\Gamma_s =4.5 \times 10^{-4}$ m$^2$/s; the prescribed vertical normal velocity component $v_z = -3.0$ mm/s and zero outside and inside the cylinder, respectively.
The cylinder drawn in each panel represents the surface of the superfluid \He{}, where the giant vortex (with circulation $\Gamma_s$) is assumed to exist.
Initially, three vortex lines exhibiting a Kelvin wave excitation are placed around the giant vortex.
The vortex lines and giant vortex are aligned mutually parallel, hence they tend to repel each other.
However, because $\Gamma_n > \Gamma_s$, the singly quantized vortex lines are pulled toward the cylinder under mutual friction.
In the presence of external flows proportional to $1/r$, the vortices are stretched and spiraled in toward the cylindrical surface, as shown in Fig. \ref{fig: Region1_VFM}.
Locally, the orientation of the vortex line near the wall is almost parallel to that of the wall; eventually, the tip of the vortex \SI{reaches} the surface.

\begin{figure}[t!] 
	\includegraphics [width=1\columnwidth]{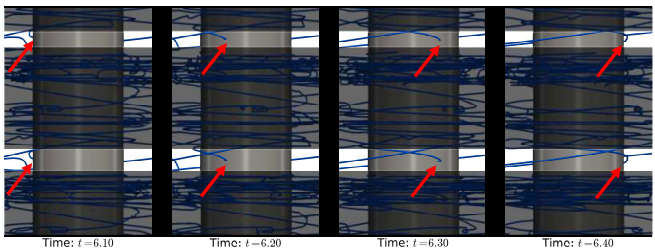}
	\caption{Series of magnified snapshots of VFM simulation at $t = 6.1$, $6.2$, $6.3$, and $6.4$ s from left to right, respectively (color online). The panels are shaded to render more clearly the growth of the helical structure, which is caused by the motion of the reconnected vortices.}
    \label{fig: Region1_VFM_2}
\end{figure}

%
In this simulation, special attention must be paid when handling the reconnection events between the singly quantized vortices and the giant vortex.
When a vortex line approaches and hits the surface of the hollow cylinder of the giant vortex, a reconnection event is highly likely; this is thought to be a crucial mechanism that sustains the growth of the circulation $\Gamma_s$ when $\Gamma_n > \Gamma_s$.
However, the conventional method of managing these events algorithmically \cite{Schwarz85} may not be valid in this system, because the boundary condition at the surface of the giant vortex is unknown.
We can assume that the singly quantized vortex lines must intersect the surface of the giant vortex perpendicularly, so that the superfluid does not flow out of the fluid through the boundary.
The perpendicularity of the vortices at the reconnecting points is approximately attained by introducing an ``effective friction'' to the ends of the vortex lines where they meet the wall of the giant vortex.
In the numerical simulation, we simply set the azimuthal and vertical velocity components of the vortex segment to be zero when it enters the cylinder through the wall.
The reconnected segments circles around the giant vortex, and the remaining vortex lines are wound around the cylinder; this can be observed in the panels in Fig \ref{fig: Region1_VFM_2} and in the video found in Ref. \cite{Sup1}.
\SI{ However, in the presence of the vertical normal flow, only the vortex segments whose orientations are such that they induces a superfluid flow along the normal flow grow selectively; meanwhile, those with the opposite orientation tend to diminish gradually through mutual friction.
This means that spiral-shaped vortex filaments with the same helical orientation are tend to be formed, which is be similar to the vortex mill discussed by Schwarz in Ref. \cite{Schwarz90}. }
%

%
Through the processes discussed in this section, quantized vortices with a specific orientation were selectively produced in Region I; they then travelled to Region II.
As vortex lines continue to wind around the giant vortex, the value of the circulation $\Gamma_s$ increases.
When the value of $\Gamma_s$ becomes sufficiently close to that of $\Gamma_n$, the giant vortex no longer attracts the free vortices, and the vortices enter a quasi-stable equilibrium state.
The vortices steadily produced in Region I can behave as a ``vortex bath,'' which is essential to bundle formation in Region II; we discuss this in Sec. \ref{sec: IV}.

\section{Bundle Formation in Region II} \label{sec: IV}
%
\SI{In the presence of a steady down-flow and an azimuthal-flow of normal fluid in Region II, some characteristic structural patterns/ polarization may be imprinted on the vortices that are densely produced in Region I and transported to Region II.
}
\SI{We consider} the normal fluid velocity profile given by Eqs. \eqref{eq: vn in} and \eqref{eq: vn out}, and we neglect the flow profile perturbation attributable to the quantized vortices generated through mutual friction.
\SI{Microscopically, this assumption does not hold. Recent studies \cite{Yui2018,Biferale19,Yui2020,Galantucci20} have shown that the normal fluid profile is non-trivially modulated by the presence of quantized vortices, through mutual friction on the scale of the inter-vortex distance.
However, in the analysis below, we only consider the macroscopic vortex bundle structure that develops in the macroscopic steady normal flow; a study of the characteristic small-scale structures that emerge due to coupled dynamics remains a future work to be dealt with. }
%

One factor that characterizes the \SI{macroscopic} vortex bundle structure is the ratio of the vertical velocity $v_z$ to the azimuthal velocity $v_\phi$ of the normal fluid.
To observe the effects of this factor, we consider a helical vortex line $\B{s}(\xi)$ with arc length parametrization $\xi \in \mathbb{R}$:
\begin{equation}
\B{s}(\xi) \equiv \left(  
\begin{array}{c} x(\xi)\\  y(\xi) \\  z(\xi)
\end{array}
\right) = \left(  \begin{array}{c}
X_0 \cos k_0 \xi\\  Y_0 \sin k_0 \xi \\  \xi
\end{array}   \right) .
\end{equation} 
On the right-hand side of Eq. \eqref{eq: EoM VFM}, we neglect all terms except the one proportional to $v_n$ (the second term); then, the equation of motion for \SI{$r<R_0$} simplifies to
\begin{equation}
\begin{split}
\B{\dot{s}}(\xi,t) & \approx \alpha \B{s}'\times \B{v}_\textrm{n} \\
& = A                                                                                                              \left(  \begin{array}{c}
\left( k_0 v_z \frac{X_0}{Y_0} - \frac{\Gamma_n}{2\pi R_0^2}\right) x \\
\left( k_0 v_z \frac{Y_0}{X_0} - \frac{\Gamma_n}{2\pi R_0^2}\right) y \\
\frac{\Gamma_n k_0}{2\pi R_0^2}  \left( \frac{Y_0}{X_0} -  \frac{X_0}{Y_0} \right)   \end{array}
\right),
\end{split} 
\end{equation}
where $ A = {\alpha}/{\sqrt{ k_0^2 (X_0^2 + Y_0^2) + 1}}$. 
When $X_0 = Y_0$, the equation of motion for the helix amplitude $r \equiv \sqrt{x^2 + y^2}$ is simply
\begin{equation} \label{eq: EoM r}
\dot{r} = \left( k_0 v_z  - \frac{\Gamma}{2\pi R_0^2}\right) r.
\end{equation}
Equation \eqref{eq: EoM r} indicates that when $\frac{\Gamma_n}{2\pi R_0^2} > k_0 v_z $, the right-hand side of Eq. \eqref{eq: EoM r} becomes negative, and the amplitude $r$ diminishes.
Assuming that the maximum wavelength of a vortex line in such a rotating normal fluid tube (radius: $R_0$) is at most $\lambda_\textrm{max} \equiv 2\pi /k_{0,\textrm{min}} \sim 2 R_0 $, then the criterion for the helical excitation on the vortex line to diminish becomes  
\begin{equation} \label{eq: ratio}
\frac{v_\phi}{v_z} \gtrsim \pi,
\end{equation}
where $v_\phi \equiv \Gamma_n / 2\pi R_0$ is the azimuthal velocity at radial distance $r = R_0$.
The validity of the criterion is confirmed through numerical simulations of the VFM in Sec. \ref{sec: IV subsec: A}.

\subsection{VFM simulations for Region II} \label{sec: IV subsec: A}
\begin{figure*}[t!] 
	\includegraphics [width=2\columnwidth]{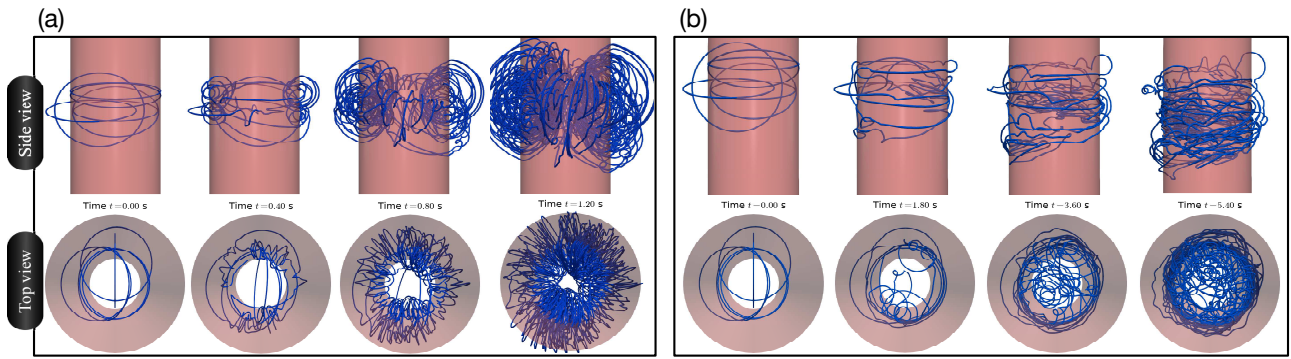}
	\caption{(a) Snapshots of VFM simulation with $v_z = 10$ mm/s and $v_\phi = 10\times \pi$ mm/s. The ratio $v_z / v_\phi$ is chosen to satisfy the relation in Eq. \eqref{eq: ratio}. The formation of a bundle of vertical vortex lines can be observed in the central region within the cylindrical shell of radius $R_0 = 2.5$ mm. (b) Snapshots of VFM simulation with $v_z = 10$ mm/s and $v_\phi = 1\times \pi$ mm/s. Because the relation is not satisfied, the amplitudes of the excitations are significant, and eventually the cylinder of radius $R_0$ is covered by helical vortex lines.}
    \label{fig: Region2_VFM}
\end{figure*}
%
We consider the dynamics of six seed vortex rings randomly placed near the central region of radius $R_0$ \SI{ (which is shown as the shaded region in Fig. \ref{fig: Schematics}(b) schematically) and see how the flow ratio modifies the polarization of the growing vortices.
}  
In numerical simulations, we set the radius $R_0 = 2.5$ mm and $v_z = -10$ mm/s, and we adjust the circulation of normal fluid $\Gamma_n$ such that $v_\phi = 10\times \pi$ and $1\times \pi$ mm/s.
Figure \ref{fig: Region2_VFM} (a) and the video in Ref. \cite{Sup2} show the case in which $v_\phi / v_z = \pi$.
Small excitations/Kelvin waves in the horizontal direction on the vortex lines are visibly damped, and straight vortex lines tend to align themselves and lengthen in the central region along the $z$-axis, as the rough estimate in Eq. \eqref{eq: ratio} indicates.
However, when the ratio was sufficiently small, the amplitudes of the Kelvin waves are amplified; this can be seen in Fig. \ref{fig: Region2_VFM} (b) and in the video found in Ref. \cite{Sup2}.
A helical excitation is amplified in the flow cylinder of radius $R_0$.
However, when the radius of the excitation exceeds $R_0$, it ceases to grow because of the absence of normal flow \SI{that} transfers energy through mutual friction.
\SI{As more helical excitations are generated, a helically polarized vortex bundle is formed.
Now, the individual vortices are repelled from the central region and form a ``vortex layer'' surrounding the cylinder of radius $R_0$.
The vortex layer induces a superfluid flow inside cylinder of the vortex layer, analogous to a magnetic field generated by a current passed through a coil.
}
%

In both cases, the growth of the vortices along the $z$-axis appears to be indefinite while the steady normal flow profile is prescribed; however, in simulations, the maximum vortex line density is limited by the computational resolution $\Delta \xi$.
Furthermore, in reality, the dense vortex bundle would significantly deform the normal profile, and our method will eventually break down.
We note that our above analysis only applies to the initially growing state of the bundle; however, it is crucial for understanding the structure of the bundle.

\SI{The growth of the vortex line density in such an external flow may be obtained in the numerical simulations in the framework of the HVBK hydrodynamics as well.
However, in the HVBK framework the quantized vortices are treated as a coarse-grained vortex-line density field in which the microscopic information of vortices, such as local curvature, is lost.
Therefore, such a method may not be suitable to investigate the vortex bundle structure directly ascribed to individual vortex dynamics.
} 

\section{Estimation of Diffusion Time Scale} \label{sec: V}
%
\SI{We have discussed the structure of the vortex bundle in Region II. 
However, the direct determination of the bundle in experiments is difficult.
Therefore, we propose that the structure (polarization/helicity) of the bundle may be assessed qualitatively by the determination of the diffusion constant of the bundle.}

The bundle in the steady state is energetically sustained by the normal fluid; thus, when the rotor is stopped, the normal flow slows down and the bundle diffuses.
The diffusion constant $D$ of a homogeneous vortex tangle is reported to be of the order of the circulation quantum number $\kappa = h/m$ \cite{Tsubota03,Nemirovskii10, Pomyalov20}.
However, in our case, the bundle is assumed to possess an ordered structure; this would allow the system to have a structure-dependent diffusion constant, which is an experimentally measurable quantity.
%

\begin{figure}[t!] 
	\includegraphics [width=1\columnwidth]{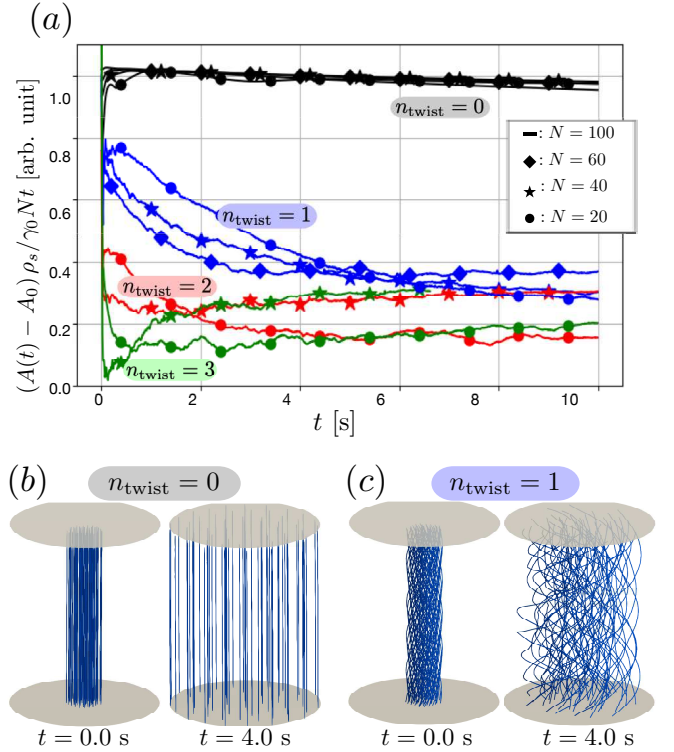}
	\caption{(a) Normalized cross-sectional areas of the bundles as functions of time for various numbers of vortices $N$ and twists $n_\textrm{twist}$. The values of the functions are proportional to the diffusion constant $D$. The proportionality constant is found in Eq. \eqref{eq: D}. (b) -- (c) Snapshots of VFM simulations with $N = 60$ vortices for $n_\textrm{twist} = 0$ and $1$, respectively. The top and bottom boundaries are subject to the periodic boundary condition. The disks in each panel are of radii $0.8$ mm. The video can be found in Ref. \cite{Sup4}}
    \label{fig: Diffusion}
\end{figure}

We consider a system of $N$ vertical, mutually parallel quantized vortices distributed evenly within a cylindrical region of radius $R_0 = 0.25$ mm.
The height of the system is set to $2.0$ mm, and the bottom and top surfaces are subject to the periodic boundary condition.
The normal fluid component is set to be stationary; thus, the vortices tend to move farther apart from each other through mutual friction.
When all the vortices are straight and perpendicular to the $z$-axis ($n_\textrm{twist}=0$) (as shown in Fig. \ref{fig: Diffusion} $(b)$), the scenario is relatively simple:
The vortices form a triangular lattice as the radius $R$ of the occupied cross-sectional area grows from its initial value $R_0$. 
Then, the superfluid velocity within the radius $R$ mimics a rigid rotation. 
However, when the bundle is ``twisted'' such that all the vortices are helically deformed (as in Fig. \ref{fig: Diffusion} $(c)$), the situation becomes more complex.

First, to qualitatively understand the diffusion process in this system, we consider the kinetic energy $E_R$ of a bundle of $N$ vertical vortex lines confined in a region of radius $R$. 
For simplicity, we assume that the vortices are not twisted (\ie{}, $n_\textrm{twist}=0$) and that the superfluid velocity profile induced by the vortices is given (in cylindrical polar coordinates) as
\begin{equation} \label{eq: vs in}
\B{v}_\textrm{s}(r,\phi,z) = \left(
\begin{array}{c} 
0\\
\frac{\Gamma_s}{2\pi}\frac{r}{R^2} \\
0
\end{array}
\right)  \quad \text{for $r<R$,}
\end{equation}
and
\begin{equation} \label{eq: vs out}
\B{v}_\textrm{s}(r,\phi,z) = \left(
\begin{array}{c}
0\\
\frac{\Gamma_s}{2\pi}\frac{1}{r} \\
0
\end{array}
\right)  \quad \text{for $r>R$},
\end{equation}
where $\Gamma_s =\kappa N$.
Then, the kinetic energy per unit height can be calculated as $E_R / L_z = (\rho_s / 2) 2 \pi \int_0^{R_\text{max}} dr r v_s^2 $.
Substituting Eqs. \eqref{eq: vs in} and \eqref{eq: vs out} into the integral, the energy is expressed as
\begin{equation} \label{eq: vs Kin Energy}
\frac{E_R}{L_z} = \frac{\Gamma^2 \rho_s}{4 \pi}    \left[   \frac{1}{4} +  \ln R_\textrm{max} - \ln R   \right],
\end{equation}
where $R_\text{max}$ is the radius of the cylindrical container.
In terms of the area $A \equiv \pi R^2$, the time derivative of Eq. \eqref{eq: vs Kin Energy} is
\begin{equation} \label{eq: time dep}
\frac{d}{dt} \frac{E_R}{L_z} =- \frac{ \Gamma^2 \rho_s }{ 8 \pi }  \frac{ \dot{A} }{A}.
\end{equation}
We can also estimate the energy dissipation rate $\varepsilon$ from the mutual friction per unit length between the resting normal fluid and the vortex lines.
In the first-order approximation, the frictional force $\B{f}$ per unit length of a vortex segment is known to be proportional to its velocity, and the proportionality constant $\gamma_0$ depends on the temperature $T$ \cite{Sup5}.
Therefore, we obtain
\begin{equation} \label{eq: Energy Dissipation}
\begin{split}
\varepsilon &= \B{f} \cdot \B{v}_s   \\
&= \frac{\gamma_0 \Gamma_s^2}{4 \pi} \sum_{i=1}^{N} \frac{r_i^2}{R^4}\\
&\approx \frac{\gamma_0 \Gamma_s^2 N}{8 \pi A}.
\end{split}
\end{equation}
The sum $\sum_{i=1}^N r_i^2$ in the second line is approximately evaluated as $NR^2/2$, assuming an even distribution.
The only major factor determining energy loss in the system is the mutual friction; thus, we equate Eq. \eqref{eq: time dep} and Eq. \eqref{eq: Energy Dissipation} to finally obtain
\begin{equation} \label{eq: Area as func of t}
A(t) = A_0 + \frac{\gamma_0 N}{\rho_s}t.
\end{equation}
Figure \ref{fig: Diffusion} $(a)$ plots the computationally obtained values for the properly normalized areas of the bundle cross-section (\ie{}, $\left( A(t) - A_0 \right)\rho_s / \gamma_0 N t $) as functions of time $t$.
It can be clearly seen that when $n=0$, the values agree with Eq. \eqref{eq: Area as func of t}.
However, they start to diverge as time elapses; thus, higher-order estimates are needed for a more precise discussion.
Interestingly, when $n_\textrm{twist}>1$, the diffusion of the bundle is strongly suppressed.
Although a clear relationship between the number of twists $n_\textrm{twist}$ and the reduction from unity in Fig. \ref{fig: Diffusion} $(a)$ has not yet been established, the significant suppression of vortex bundle diffusion can be expected in the experiments if the bundle is twisted.

%
The expression in Eq. \eqref{eq: Area as func of t} relates to the diffusion constant $D$ in conventional $2$-dimensional diffusion problems; that is, $\dot{n} = D \nabla^2 n$.
A solution to the partial differential equation, using an instantaneous delta function-like source at time $t=0$, takes the form
\begin{equation}\label{eq: Soln to DE}
n(r,t) = \frac{N}{4\pi D t} \exp \left(- \frac{r^2}{4Dt}  \right),
\end{equation}
where $n(r,t)$ is the vortex number density such that $N = 2\pi \int_0^\infty r n(r,t) dr $ is the total number of vortices.
The radius $R$ of the cross-sectional area of the bundle is characterized by the exponential function in Eq. \eqref{eq: Soln to DE}, and $R \sim \sqrt{4Dt}$.
Combining this result with Eq. \eqref{eq: Area as func of t}, we obtain the final expression:
\begin{equation} \label{eq: D}
D \approx \frac{\gamma_0 N}{4 \pi \rho_s}.
\end{equation}
In the experiment at OCU, because the temperature $T$ was $1.6$ K and the number of vortices $N$ was of order $10^4$, the diffusion constant was approximately $D \approx 8$ mm$^2$/s.
The values of the temperature-dependent quantities $\gamma_0$ and $\rho_s$ can be found in Ref. \cite{Barenghi83}.
%

%
If we linearly extrapolate our computational results for the simplified system, then the diffusion constant measurable in the experiment is $\sim 2$ mm$^2$/s if the vortex bundle is twisted.
In our above analysis, the normal fluid is assumed to be at rest for the sake of simplicity; however, in the case of an experiment where $10^4$ vortices are present, this assumption may not be valid.
If the bundle of vortices and the normal flow co-rotate about the $z$-axis, then the energy loss via mutual friction in Eq. \eqref{eq: Energy Dissipation} is reduced; this would lead to further reduction of the diffusion constant\SI{, at least initially.
Therefore, we would need to wait for sufficiently long time after the rotor is stopped (so that the vortex line density becomes small and normal fluid comes to rest) to observe the predicting decay behavior. }

\section{Conclusions and Discussion} \label{sec: VI}
%
Motivated by an experimental report on the ``bathtub'' vortex of superfluid \He{}, we discussed the structure of the quantized vortex bundle that can be formed in such a macroscopic flow, based on numerical simulations using the VFM.
The superfluid bathtub vortex system was investigated by separating it into three regions.
The top region (Region I) is assumed to contain a giant vortex with multiply quantized circulation.
\SI{By analogy with rotating superfluid \He{}, we illustrated the development process of the giant vortex (or the surface dimple). }
In Region I, a vortex bundle can develop alongside the giant vortex.
\SI{The bundle that forms around the giant vortex appears to act as a major source of the vortices that are transferred to Region II; thus, it can be considered as a ``vortex-line bath.''  }
Region II is the region in which the boundary effect of the vessel bottom is negligible and a vortex-line bath is present \SI{at the top}.
Because the normal fluid has an intrinsic viscosity, we assume that it establishes a macroscopic steady flow.
The steady normal flow ``stirs'' the transferred vortex loops; this presumably deforms the bundle structurally, reflecting the geometry of the normal flow.
Then, the bundle settles in a steady state such that the mutual friction between the two fluids is minimized.
%

The velocity profile of the normal fluid in our analysis is that of a Rankine-vortex-like flow, containing a vertical flow within a radius $R_0$ along the $z$-axis, as described in Eqs. \eqref{eq: vn in} and \eqref{eq: vn out}.
In such environments, the vortices that constitute a bundle either (a) align themselves parallelly along the $z$-axis or (b) wind around the down-flow region of radius $R_0$ and form a cylindrical vortex layer.
Whether the bundle takes the structure (a) or (b) depends on the ratio of the vertical velocity $v_z$ to the azimuthal velocity $v_\phi$ of the normal fluid.
Because of the complexity of the experimental setup, no direct experimental data are currently available to indicate size of the ratio.
Instead of measuring the ratio, we proposed that the structure could be elucidated indirectly, by measuring the decay of the vortex bundle.
In the OCU experiment, the expected vortex diffusion constant $D$ was approximately $8$ mm$^2$/s, if the bundle was not twisted along the $z$-axis.
A series of VFM simulations indicate that the diffusion constant is significantly reduced if the bundle is twisted.
By experimentally measuring the extent to which the diffusion constant diverges from its expected value, we can further our understanding of the structures of vortex bundles in macroscopic bathtub vortices.

\begin{acknowledgments}
We thank Ken Obara for useful discussions.
This work was supported by JSPS KAKENHI Grant No. JP20H01855. 
S. I. was supported by Grant-in-Aid for JSPS Fellow Grant No. JP20J23131.
\end{acknowledgments}

\bibliography{SV}

\begin{thebibliography}{41}
\bibitem [{Sup({\natexlab{c}})}]{Sup5}%
  \BibitemOpen
  \href@noop {} {} ({\natexlab{c}}),\ \bibinfo {note} {the friction force
  $\bm{f}$ on a vertical vortex filament that moves with $\bm{v}_L$, the
  velocity in the laboratory frame, per unit length is expressed as $ \bm{f} =
  \gamma_0 (\bm{v}_n - \bm{v}_L) + \gamma_0 ^\prime\bm{s}^\prime \times
  (\bm{v}_n - \bm{v}_L ). $ The relation between this friction parameter
  $\gamma_0$ and the mutual friction parameters, $\alpha$ and $\alpha ^\prime$,
  discussed in Eq. $(2)$ is given as $$ \gamma_0 = \frac{\rho_n \rho_s}{2\rho}
  \kappa \frac{\alpha}{(1-\alpha' \rho_n /2\rho)^2 + \alpha^2 \rho_n^2
  /4\rho^2}$$ and $$ \gamma_0' = \frac{\rho_n \rho_s}{2\rho} \kappa
  \frac{\alpha^2 \rho_n /2\rho - \alpha' (1-\alpha' \rho_n /2\rho)}{(1-\alpha'
  \rho_n /2\rho)^2 + \alpha^2 \rho_n^2 /4\rho^2}$$
  \cite{Donnelly91,Barenghi83}.}\BibitemShut {Stop}%
\end{thebibliography}

\end{document}